\documentclass[twocolumn,english,aps,prl]{revtex4}
\usepackage{color}
\usepackage{amsmath}
\usepackage{graphicx}

\definecolor{dgreen}{rgb}{0.0,0.5,0.0}

\usepackage[normalem]{ulem}
\renewcommand{\Im}{\mbox{Im}} 
\newcommand{\nag}{{\phantom{\dagger}}}

\usepackage{babel}

\newcommand{\beq}{\begin{equation}}
\newcommand{\eeq}{\end{equation}}
\newcommand{\beqn}{\begin{eqnarray}}
\newcommand{\eeqn}{\end{eqnarray}}
\newcommand{\bgg}{\begin{gather*}}

\newcommand{\gweisso}{{\cal G}_0}
\newcommand{\kk}{\mathbf{k}_{\parallel}}

\global\long\def\eps{\varepsilon}
 \global\long\def\und#1{\underline{#1}}
 \global\long\def\volt{\Phi}
\global\long\def\dag{\dagger}

\newcommand{\vv}{\mathbf}

\newcommand{\up}{\ensuremath{\uparrow}}
\newcommand{\dw}{\ensuremath{\downarrow}}

\newcommand{\para}{\ensuremath{\parallel}}

\global\long\def\li{{\cal L}}
\global\long\def\tr{\mbox{tr}}

\newcommand{\Ree}{{\rm Re\:}}

\newcommand{\ugloc}{\und G_{\text{LOC}}}

\newcommand{\kkom}{\kk,\omega}

\newcommand{\intdkk}{\int_{BZ} \frac{d\kk}{(2\pi)^2}\:}

\long\def\wtil#1{\widetilde{#1}}
\long\def\vvwtil#1{\widetilde{\vv #1}}
\global\long\def\up{\uparrow}
\global\long\def\down{\downarrow}
\global\long\def\mdot#1{\overset{{\bf .}}{#1}}
\global\long\def\bra#1{\left<\:#1\:\right| }
\global\long\def\ket#1{\left|\:#1\:\right> }
\global\long\def\braket#1{\left<\,#1\,\right>}

\begin{document}

\title{Nonequilibrium Dynamical Mean Field Theory:
\\
 an auxiliary Quantum Master Equation approach }

\author{Enrico Arrigoni}

\email{arrigoni@tugraz.at}

\selectlanguage{english}%

\affiliation{Institute of Theoretical and Computational Physics, Graz University
of Technology, 8010 Graz, Austria}

\author{Michael Knap}

\affiliation{Institute of Theoretical and Computational Physics, Graz University
of Technology, 8010 Graz, Austria}

\author{Wolfgang von der Linden}

\affiliation{Institute of Theoretical and Computational Physics, Graz University
of Technology, 8010 Graz, Austria}

\pacs{
71.27.+a 
47.70.Nd 
73.40.-c  
05.60.Gg 
}

\begin{abstract}
We introduce a versatile method to compute electronic steady state properties 
of strongly correlated extended quantum systems out of
equilibrium. The approach is based on  dynamical mean-field theory (DMFT), in which the original 
system is mapped onto an auxiliary non-equilibrium impurity problem imbedded in a Markovian environment. 
The steady state Green's function of the auxiliary system is
solved by full diagonalization of the corresponding Lindblad equation.
The approach can be regarded as the nontrivial extension of the exact-diagonalization
based DMFT  to the non-equilibrium case.
As a first application, we consider an interacting Hubbard layer attached to two metallic leads
and present results for the steady-state current and  the non-equilibrium density of states.
\end{abstract}

\ifx\clength\undefined
\maketitle
\else
\nocomm
\fi

Due to the progress made in microscopically controlling quantum mechanical systems
within quantum optics and ultracold quantum
gases~\cite{ra.sa.97,ja.br.98,gr.ma.02,ha.br.08},
in solid
state nanoscience, spintronics, molecular electronics,~\cite{zu.fa.04,bo.gr.05}
 as well as ultrafast {laser spectroscopy~\cite{iwai_ultrafast_2003,cavalleri_evidence_2004,pe.lo.06,fausti_light-induced_2011}},
the interest in correlated systems out of equilibrium has steadily increased in
recent years. 
These achievements have prompted new and boosted old related 
theoretical questions such as nonequilibrium quantum phase transitions~\cite{mi.ta.06},
 dissipation and decoherence~\cite{le.ch.87}, and thermalization after a
 quantum quench~\cite{caza.06,ca.ca.07,ri.du.08}. 

{In this respect, the theoretical description and understanding of strongly correlated quantum systems out of equilibrium poses
an exciting challenge to modern theoretical physics.}
A widely used and successful method to treat strongly correlated
lattice systems {\it in equilibrium} 
is dynamical mean-field-theory~\cite{ge.ko.96,voll.10,me.vo.89} (DMFT).
The success of the method lies on the one hand in the nontrivial
treatment of dynamical properties, and on the other hand in its
applicability to a range of different problems, from solid states
fermionic systems to ultracold bosonic atoms, as well as the
possibility to combine it with realistic electronic structure methods~\cite{an.po.97}.
Recently, DMFT has been extended
to deal with time dependent nonequilibrium
problems~\cite{sc.mo.02u,fr.tu.06,free.08,jo.fr.08,ec.ko.09,okam.07}. 
The extensions are based on the Kadanoff-Baym-Keldysh nonequilibrium
Green's function approach~\cite{schw.61,keld.65,kad.baym,ra.sm.86}.

DMFT relies on the solution of a correlated impurity problem, which
constitutes the bottleneck of the approach.
 Several
 techniques have been adopted in the equilibrium case.
Most of them have been applied, in a more or less approximate or
limited way,
to nonequilibrium DMFT as well, either in steady state or within full
time dependence. Among them are 
iterated perturbation
theory~\cite{sc.mo.02u}, numerical renormalization group~\cite{jo.fr.08},
continuous time quantum monte carlo (CTQMC)~\cite{ec.ko.09,ec.ko.10},
noncrossing approximation and beyond~\cite{okam.08}. {Additionally, exact 
DMFT solutions are available in certain limits~\cite{fr.tu.06,tsuji_correlated_2008,tsuji_nonequilibrium_2009}}. 
{
Nonequilibrium quantum impurity problems (not within DMFT)
have also been studied by means of
 scattering-states approaches~\cite{me.an.06,ande.08},
perturbative methods~\cite{me.wi.92,sc.sc.94} in
combination with renormalization group (RG)~\cite{ro.pa.05,scho.09},
time-dependent density-matrix RG ~\cite{wh.fe.04,da.ko.04} and numerical RG~\cite{an.sc.05}
flow equation~\cite{kehr.05},  functional RG~\cite{ge.pr.07,ja.me.07},
dual fermions~\cite{ju.li.12}, and finally
CTQMC on an auxiliary system with an imaginary bias~\cite{han_quantum_2006,han_imaginary-time_2007,di.we.10}.
}

\begin{figure}
\includegraphics[width=\columnwidth]{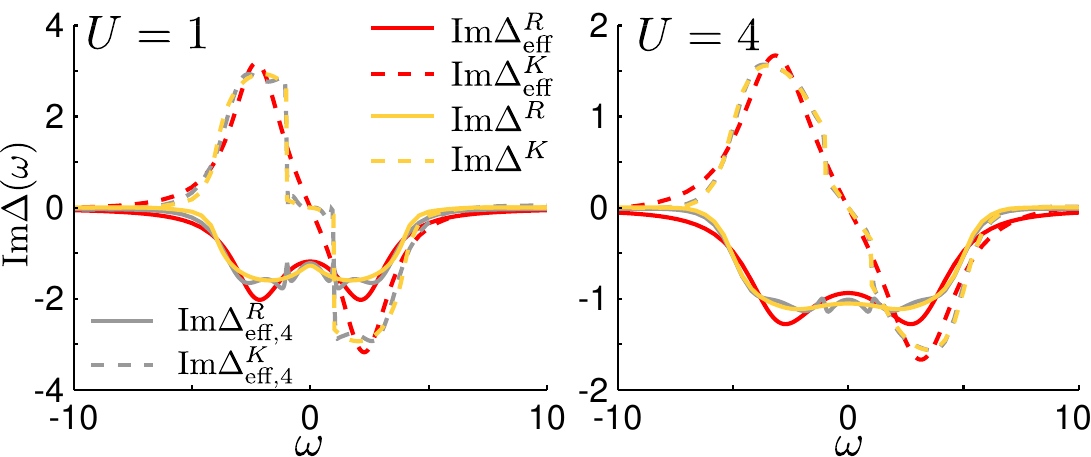}
\caption{(Color online)
Imaginary part of the 
DMFT and effective 
bath hybridization functions (retarded (R) and 
Keldysh (K) components)
for $\volt=2$, $v^2=0.1$, and two 
different values of $U$.
We also plot the effective bath hybridization function  ($\und
\Delta_{\text{eff},4}$)
 obtained 
with $N_b=4$.
\label{del}
}
\end{figure}

In this Letter, we  {
propose an approach that, 
in contrast to previous work,
while directly accessing 
steady-state properties, 
 features a solution of the  DMFT impurity problem
 {\it with controlled accuracy}. This means that the accuracy
can be directly estimated 
by comparing the exact and the
effective bath hybridization functions (Fig.~\ref{del}).
Also, no often unreliable analytical continuation is required.
}
{
At the heart of the method lies 
a solution of the
{nonequilibrium} DMFT impurity problem, 
which can be seen as a
generalization of the exact diagonalization (ED) approach, widely
used in the equilibrium case~\cite{ge.ko.96}.
However, a crucial difference with respect to conventional DMFT-ED
 is the fact that  
here the effective impurity model 
}
describes an infinite system and, thus, displays a
continuous spectrum.

In ED-based equilibrium DMFT~\cite{ge.ko.96}
a certain number of noninteracting bath sites is introduced in order
to fit the
bath hybridization function 
required by the the self-consistency
condition. The maximum number of bath sites is limited by the exponential
increase of 
 the many-body Hilbert space. 
In equilibrium, the fit is carried out in 
imaginary (Matsubara) frequency space, 
where functions are smooth,
in contrast to real frequency. There are several difficulties in this
approach when trying to extend it to nonequilibrium steady states. 
(i) Due to the finite number of bath sites, a stationary solution of the
impurity problem will always produce some equilibrium self
energy. Besides the fact that this may be questionable, 
 it is not clear which chemical potential or temperature should be
used for the impurity problem.
(ii) Secondly, due to the finite system, the bath spectrum 
is discontinuous, so that a fit in real frequencies becomes
problematic.
Unfortunately, there is 
 no Matsubara Green's function in nonequilibrium, so that this poses a
 serious problem.

The alternative presented in this Letter consists in replacing the DMFT
impurity Hamiltonian 
with an effective one
 which
is solvable by ED but at the same time  describes
a truly infinite system. 
This is obtained by connecting
the 
interacting impurity  to a moderate number of bath sites which, in turn,
are attached to {Markovian reservoirs, } 
see 
below
 for details.
{The exact bath spectral function is smoothly obtained in the
  (ideal) limit of an infinite number
of bath sites. 
}
 The action of {such} 
Markovian {baths} 
on the reduced density matrix of the
system consisting of the other bath sites and of the impurity, 
is described by
the Lindblad quantum Master equation~\cite{br.pe}.
The latter  can
be readily solved by 
diagonalizing the Lindbladian within the many-body ``super-Fock''
space of reduced density matrices of the system. %
Its solution determines both the retarded and Keldysh impurity 
Green's function, as well as the self-energy. 
The latter is used in the DMFT loop to obtain the new 
bath hybridization function, which 
is fitted by new bath parameters.

\begin{figure}
 \includegraphics[width=\columnwidth]{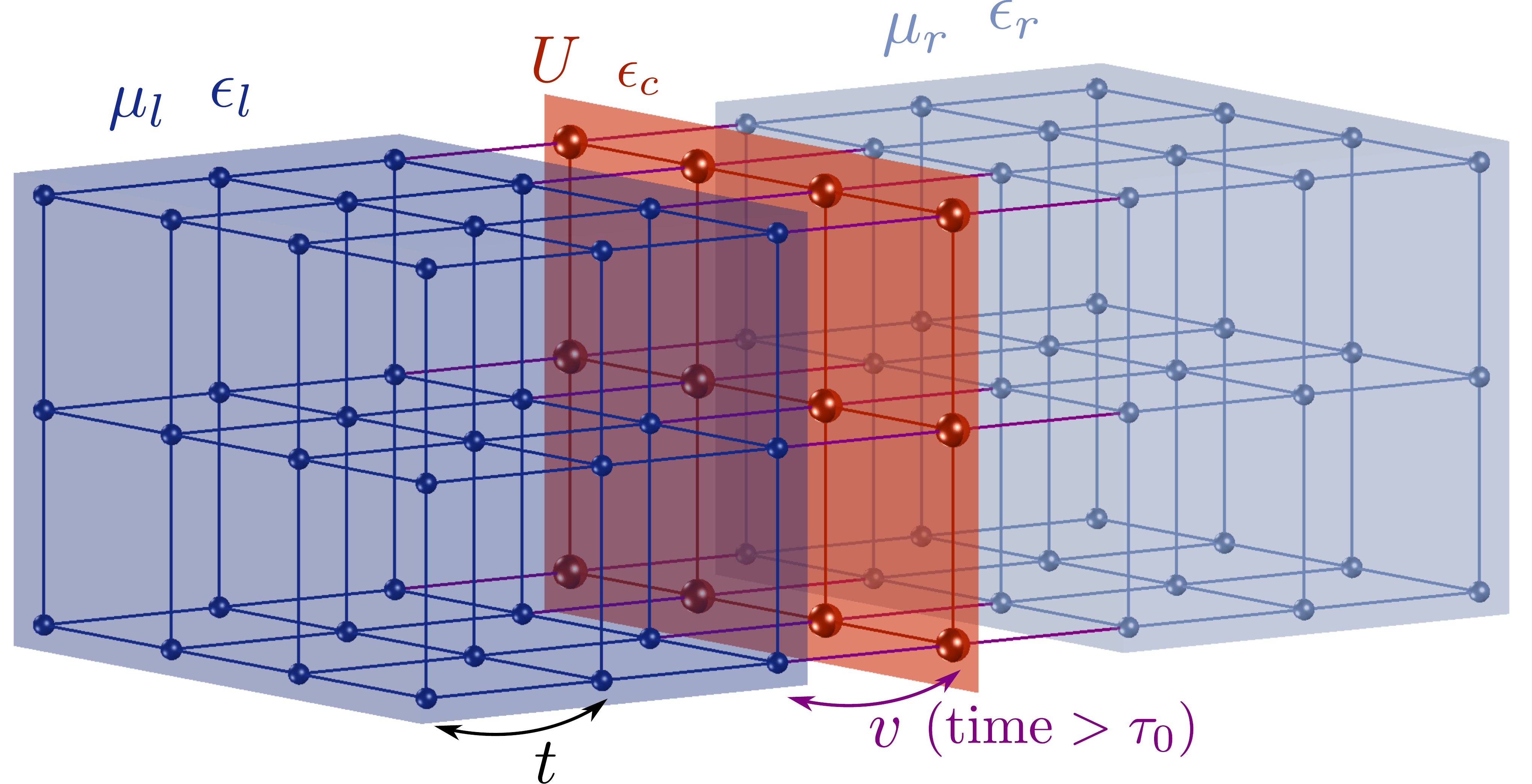}
\caption{(Color online)
Schematic representation of the system at study.
\label{schematicp}
}
\label{file}
\end{figure}

In order to illustrate the approach, we apply it
 to a simple model describing a 
heterojunction consisting of a correlated
interface {$(c)$ sandwiched} between two
 metallic leads {$\alpha=l,r$ (see Fig.~\ref{schematicp})}.
{Experimentally, such a setup has been recently explored where the 
correlated layer was realized by a V${}_2$O${}_3$ microfilm that is coupled to
Au leads~\cite{guenon_electrical_2012}.}

Before 
 a certain time $\tau<\tau_0$ the three regions {$c,\,l,\,r$} are 
disconnected and in equilibrium 
 at different chemical potentials $\mu_c, \mu_l$, and $\mu_r$, respectively.
This  
{amounts to} applying 
a bias voltage 
$\volt=\mu_l-\mu_r$ between the leads.
The central
region, lying on the $x-y$ plane, is described by a single-band Hubbard layer on a square
lattice   with onsite interaction $U$, onsite 
energy $\eps_{c}$, and nearest-neighbor hopping $t$. 
The leads consist of
two half-infinite cubic lattices described by a nearest-neighbor noninteracting
tight-binding model with hopping 
which we take as unit of the energy
$t=1$, 
and on-site energies $\eps_\alpha$.
  We restrict for simplicity
to the particle-hole symmetric case for which 
$\eps_{c}=-U/2$,  $\mu_r=-\mu_l$,  and $\eps_r=-\eps_l$. Finally, we
take $\mu_\alpha=\eps_\alpha$, which corresponds to having the same
electron densities in the two leads. 
A related nonequilibrium model has been treated in DMFT within
perturbative impurity solvers in Refs.~\cite{okam.07,okam.08}.

Starting at $\tau=\tau_0$, 
 a nearest-neighbor
hopping $v$ is switched on between the central region and the leads. 
After a sufficiently long time, a steady-state
is reached, provided no trapped surface states occur. 
Nonequilibrium properties, in general, and nonlinear transport in
particular can quite generally be addressed in the framework of the Keldysh
Green's function approach~\cite{kad.baym,schw.61,keld.65,ha.ja,ra.sm.86}.
Here, we adopt the notation (see e.g. Ref.~\cite{ra.sm.86}) 
{where} the
(underlined) Keldysh Green's function is a $2\times2$ 
matrix containing 
the retarded ($G^{R}$), advanced ($G^{A}$), and Keldysh ($G^{K}$)
components. In steady state, these depend on a single frequency only.
 The system is translation invariant in the direction parallel to the
layer, which we denote as $\para$. Accordingly we can write Dyson's
equation for the layer (c) Green's function $\und G(\kkom)$, 
$\kk$ being the $\para$ momentum~\cite{ha.ja} 
as
\beq
\label{gkkom}
 \und G(\kkom)^{-1} = 
 \und g_0^{-1}(\omega) - \sum_{\alpha=l,r} v^2
  \und g_\alpha(\kkom) -
\und\Sigma(\kkom)
\;.
\eeq
Here, $\und\Sigma(\kkom)$ is the self-energy,
$\und g_0$ is the 
$v=0,U=0$ layer Green's function, and
$g_\alpha(\kkom)$ are the $v=0$ leads Green's functions 
on the first lead layers. Their retarded and Keldysh components
are readily obtained analytically in terms of 
the Green's function of a half-infinite tight-binding
chain.

Within DMFT, one  approximates the self energy by a local,
i.e. $\kk$-independent  $\und\Sigma(\omega)$, which is determined
by solving a quantum impurity model  with the same interaction $U$ 
embedded in a self-consistently
determined bath~\cite{ge.ko.96}.
The latter is completely specified
by the bath hybridization function $\und \Delta(\omega)$, which
is determined
  self-consistently by requiring that the Green's
function of the impurity $\und G_{\text{IMP}}(\omega)=\left(\und 
  \gweisso(\omega)^{-1}-\und\Sigma(\omega)\right)^{-1}$ be equal to
the local Green's function of the layer (cf.~\cite{ge.ko.96,fr.tu.06,okam.07}) 
$
\ugloc(\omega) = \int \frac{d\kk}{(2\pi)^2}\ \und G(\kkom)  
$, 
{where} $\und G(\kkom)$ is given by \eqref{gkkom} with $\und
\Sigma(\kkom)=\und \Sigma(\omega)$ as obtained by the solution of the
impurity problem.
Here, $\und \gweisso$ is 
the ``Weiss'' bare Green's function of the impurity model 
defined as
$
\und \gweisso(\omega)^{-1}=\und g_0(\omega)^{-1}-\und\Delta(\omega)
$.

The  solution of the
impurity problem is in fact the DMFT bottleneck. 
The usual (equilibrium) DMFT ED procedure consists
in approximating the effect of the total bath hybridization function $\und\Delta(\omega)$
by an ``effective'' bath 
{with} a finite number $N_b$ of bath sites.
Quite generally, {in equilibrium} one carries out some fit to the bath hybridization
function in Matsubara space.
As discussed 
above,
 this is not appropriate in nonequilibrium
steady state. In this Letter, we present a different approach: In additional
to a certain (even) number $N_b$ of bath sites, which we more conveniently connect
to the impurity in the form of two chain segments, 
 we include two Markovian baths, which
represent a particle reservoir and sink, respectively. 
{Their role is to compensate }
for the ``missing'' part of the
infinite chain which would be necessary to exactly reproduce the desired
$\und\Delta(\omega)$.
 The bath parameters, i.e. hopping and on-site energies
of the bath sites, as well as the Lindblad coefficients (see below) of
the Markovian baths, are then fitted to $\und\Delta(\omega)$. More specifically,
we minimize the cost function 
$
\int d\omega\sum\limits_{x=R,K}
\left(\Im(\Delta^{x}(\omega)-\Delta_{\text{eff}}^{x}(\omega))\right)^{2}
$,
where $\und\Delta(\omega)$ is 
obtained from 
$\ugloc$ 
 via
\beq
\label{gloci}
\und\Delta(\omega) =
\und g_0(\omega)^{-1}-\ugloc(\omega)^{-1} - \und\Sigma(\omega) \;,
\eeq
while
 $\und\Delta_{\text{eff}}$ is the 
bath hybridization function produced
by the effective bath (bath sites+Markovian baths).
An important aspect
is that, although the (outermost) baths are Markovian, their effect
on the impurity site is non Markovian due to the presence of the intermediate
bath sites. This can be seen, for example, in the spectrum of
$\und\Delta_{\text{eff}}$ in Fig.~\ref{del}, 
 which in the Markovian case would be a constant.
Furthermore, upon increasing the number $N_b$ of intermediate
bath sites, the effect of the Markovian bath becomes weaker 
and one is expected to approach the exact result
$\und\Delta_{\text{eff}}(\omega)=\und\Delta(\omega)$ for large $N_b$. 

We now specify the effective bath more in detail. This consists of
an Hamiltonian for the ``system'' (a chain of impurity+bath sites)
\begin{equation}
H=\sum\limits _{n, m,\sigma}E_{n,m}c_{n\sigma}^{\dag}c_{m\sigma}^\nag+
U
c_{0\up}^{\dag}c_{0\up}^\nag c_{0\dw}^{\dag}c_{0\dw}^\nag, 
\end{equation}
in usual notation, where $0$ is the impurity site, and $n=-1,\cdots,-l$
($n=1,\cdots,l$) are left (right) bath sites. Here, {$E_{0,0}=\eps_{c}$},
and the bath energies $E_{n,n},\: n\not=0$ as well as the hoppings 
$E_{n,m},n\not=m$ are fit parameters~\cite{e00too},
whereby one can restrict to nearest-neighbors $E_{n,n\pm1}$. The
effect of the Markovian baths is expressed in terms of the Lindblad
quantum master equation which controls the time ($\tau$) dependence
of the reduced density matrix $\rho$ of the system~\cite{br.pe}:
$
\frac{d}{d\tau}\rho=\li\rho \label{rho}
$,
where $\li=\li_{H}+\li_{b}$, and $\li_{H}\rho=-i[H,\rho]$. 
{The dissipator $\li_{b}$ has the }
form
\bgg
\li_b\rho\equiv2\sum_{n,m}\Bigl(\Gamma_{n,m}^{(1)}\left(c_{n\sigma}^\nag\rho c_{m\sigma}^{\dag}-\frac{1}{2}\{\rho,c_{m\sigma}^{\dag}c_{n\sigma}^\nag\}\right)+\\
+\Gamma_{n,m}^{(2)}\left(c_{n\sigma}^{\dag}\rho c_{m\sigma}^\nag-\frac{1}{2}\{\rho,c_{m\sigma}^\nag c_{n\sigma}^{\dag}\}\right)\Bigr)\:, 
\end{gather*}
with real, symmetric, and positive definite Lindblad matrices $\Gamma_{n,m}^{(i)}$.
At first sight, one would assume a ``source'' bath 
attached to the 
leftmost and
a ``sink'' bath to the rightmost site. 
However,
the accuracy improves considerably if one allows all $\Gamma_{n,m}^{(i)}$
to be nonzero parameters to fit $\und\Delta$.

\begin{figure}
\includegraphics[width=\columnwidth]{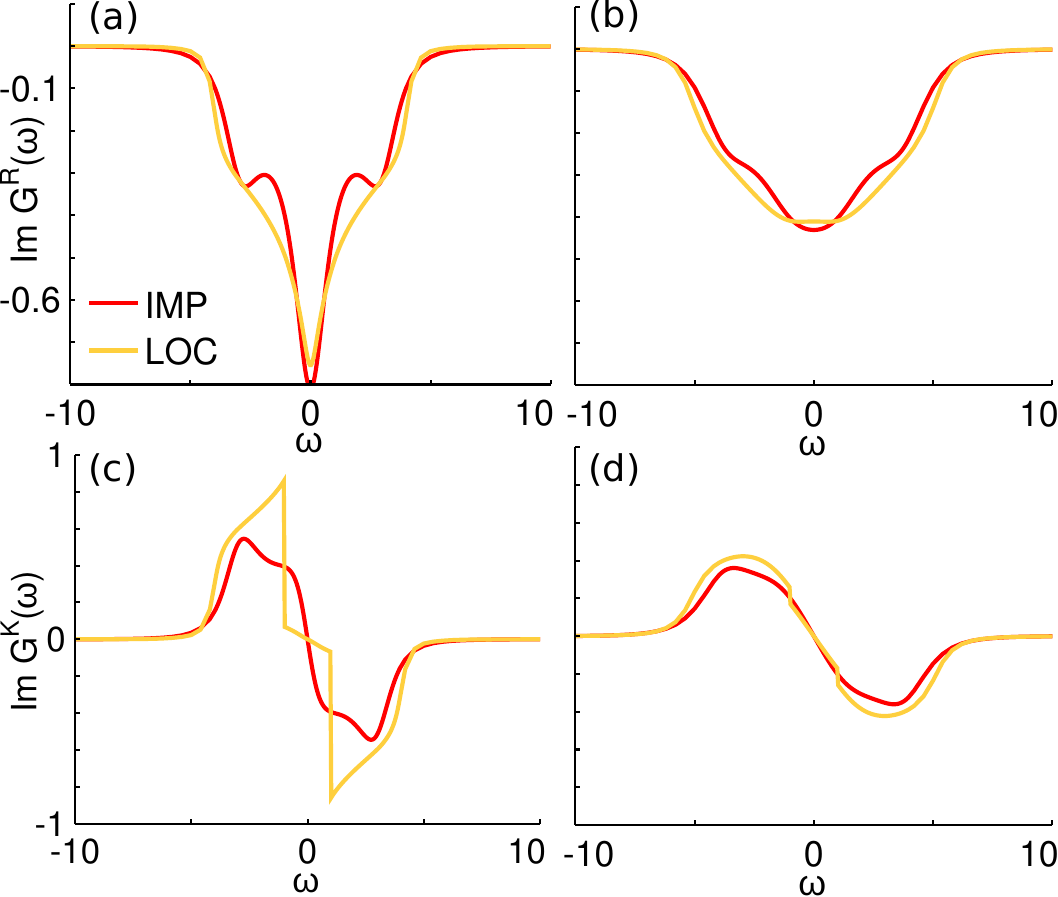}
\caption{(Color online)
Imaginary part of the impurity and local Green's function for $U=1$ ({left}) and $U=4$ ({right})
and $\volt=2$. 
\label{gru}
}
\end{figure}

As discussed in detail in Refs.~\cite{schm.78,ha.mu.08,dz.ko.11,pros.08},
the open-system problem 
describing the effective bath
 can be mapped onto a superhamiltonian
acting on a superfermion space with twice as many degrees of freedom
(i.e., ``orbitals'').
This many-body superhamiltonian, corresponding to $i\li$, which is
non-hermitian, can be diagonalized by conventional methods within the
super-Hilbert space. 
Quite
generally, 
 $\li$ has a unique eigenvector
with eigenvalue $0$, which corresponds to the steady-state density
matrix $\rho_{SS}$. All other eigenvalues have a negative real part,
corresponding to decaying terms. With the same formalism, and exploiting
the quantum regression theorem~\cite{carmichael1},
 one can evaluate correlation functions
$C_{AB}(\tau)=\tr A(\tau)B(0)\rho_{SS}$ of any pair
of system operators $A$ and $B$, and thus the required
impurity self energy $\und\Sigma(\omega)$.~\cite{supp}

The noninteracting Green's function for
the effective system+bath is necessary in order to extract
$\und{\Sigma}(\omega)$
and the bath hybridization function $\und{\Delta}_{\text{eff}}$. 
 This can be easily obtained \cite{alt}
by observing that the Markovian baths can be exactly represented by
two noninteracting fermionic baths in the wide-band limit with chemical
potentials $\pm\infty$. By taking into account the relation between
the matrices $\Gamma$ and the parameters of this bath \cite{details},
one obtains for the noninteracting system Green's function (boldface
object represent matrices 
with indices corresponding to system sites $n$):
$
\left(\vv G_{0}^{-1}\right)^R=\omega \vv I - \vv E+i(\vv
\Gamma^{(1)}+\vv \Gamma^{(2)})$,
$\left(\vv G_{0}^{-1}\right)^K=2i(\vv \Gamma^{(1)}- \vv \Gamma^{(2)})
$.

The DMFT self-consistency loop consists in (i) starting from some
initial values of the variational parameters $E_{n,m}$ and $\Gamma_{n,m}^{(i)}$,
(ii) solving the impurity problem via the approach described above
and determining $\und{\Sigma}$, (iii) evaluating
$\ugloc(\omega)$  
and 
$\und{\Delta}(\omega)$
(from \eqref{gloci})
(iv) 
determining new values of the parameters $E_{n,m}$ and
$\Gamma_{n,m}^{(i)}$ by 
minimizing the cost function, and finally
 (v) using these new parameters to
repeat the procedure from (i)
until the parameter values converge. 
Of course there is an intrinsic
inaccuracy which, for a fixed number of bath sites cannot be reduced, and
is due to the error in the fit of $\und{\Delta}(\omega)$ by a finite
number of parameters.
In principle,
this
can be systematically improved by increasing the number of bath sites.
Of course, this is limited by the exponential increase of the Hilbert
space, which, in this case, is even faster due to the fact that
the number of degrees of freedom of the superfermion space is twice
the one of the fermion space, so this makes the effort more difficult
than in ordinary ED for the same number of bath sites.
One should, however, observe that the number of fit parameters 
{is larger than} 
in the case with
``simple'' ED without Markovian bath~\cite{roug}.
\begin{figure}
 \includegraphics[width=\columnwidth]{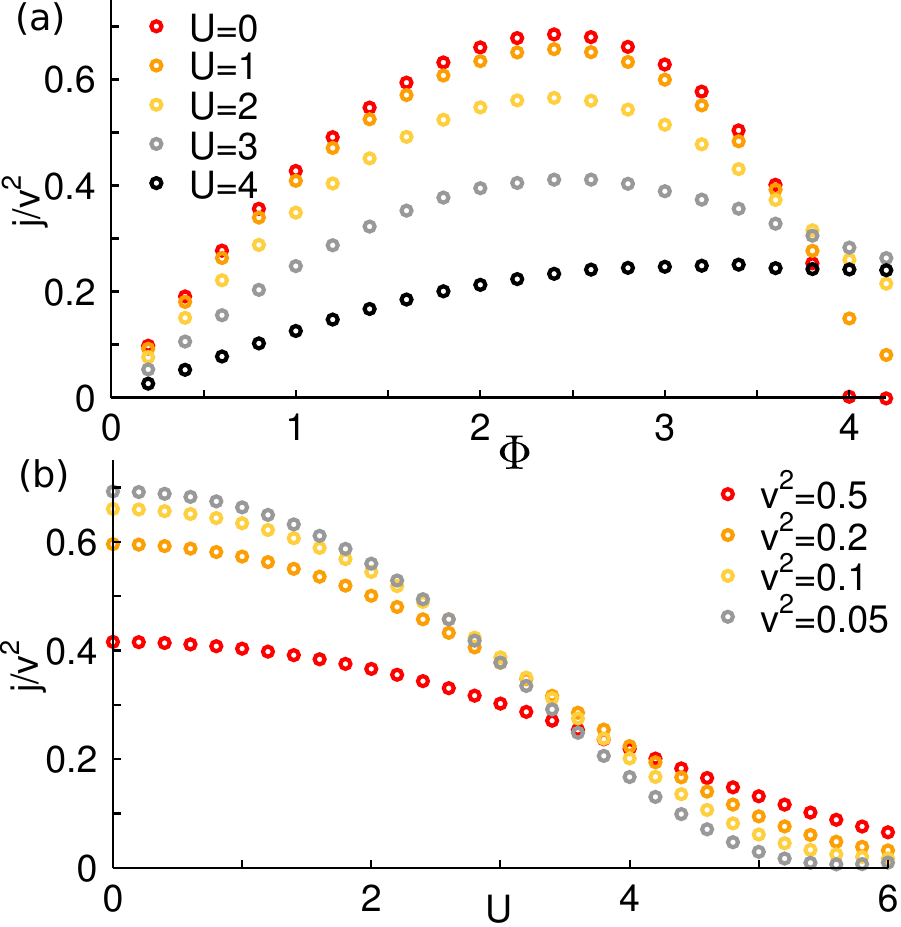} 
\caption{(Color online)
Scaled current density $j/v^2$ as a function of (a) the bias voltage $\volt$ for different values of $U$ and $v^2=0.1$, and (b)
the interaction $U$, for different values of the hybridization $v$, and $\volt=2$.
\label{jv}
}
\end{figure}

We have used
here an effective bath containing $N_b=2$  sites. Still, by
taking 
all possible parameters into account, i.e., allowing, for
example, all matrix elements of the $\Gamma$ matrices to be nonzero
(within the constraints imposed by symmetries), this gives $8$
independent fitting
parameters. In ``simple'' ED, one would have only
$2$ in the particle-hole symmetric case~\cite{para}.
We take model parameters~\cite{not} $v^{2}=0.1$, $t=1$, and 
  the leads are fixed to zero temperature.
In Fig.~\ref{del}, we show the result of the fit to the bath hybridization function for
two values of $U$ and bias voltage $\volt=2$. 
As one can see, already with this small number of bath sites, the fit
is quite good. 
Moreover, the structure of
$\und\Delta_{\text{eff}}$ is clearly 
 non-Markovian, as expected.
For comparison, we also plot the result of the fit to the bath
hybridization function obtained  with $N_b=4$ bath sites. 
This shows a considerable improvement.
The quality of the fit can be also inferred by directly
plotting the 
local and the
impurity Green's function in Fig.~\ref{gru} 
for two
different values of $U$ and $\volt=2$.

The bias voltage induces a steady-state current. The expression for
the current density $j$ (current per square plaquette) is obtained straightforwardly in terms of the
layer Green's function and the $v=0$ lead Green's functions (see,
e.g. \cite{ha.ja}) as:
\begin{gather*}
 j = v^2 \intdkk \int \frac{d\omega}{2\pi} \: \Ree \bigl[ G^{R}(\kkom)
 g_l^{K}(\kkom) \nonumber \\
+ 
 G^{K}(\kkom) g_{l}^{A}(\kkom) \bigr] \;.
\end{gather*}
This is plotted in Fig.~\ref{jv} as a function of bias voltage.
The current, as expected, decreases with increasing $U$ for smaller
biases. At larger $\volt$ the behavior is opposite, since $j$ extends
over a range of voltages which increases with increasing $U$. 
While at $U=0$, a particle going through the interface 
conserves $\kk$, and thus the current goes
to zero at a bias voltage equal to the one-dimensional ($z${-direction}) bandwidth, 
the scattering at nonzero $U$ mixes $\kk$ and thus
 broadens the bandwidth of
possible final states.
In Fig.~\ref{jv} we plot the scaled current $j/v^2$ as a function of bias voltage for
different values of $U$, as the conductance is expected to behave as
$v^2$ in a conductor, while it is suppressed ($\propto v^4$)
in a gapped system. 
The crossing  of the curves 
around $U\sim 4$ is  a signal of the nearby
equilibrium metal-insulator transition.
 However, this should be seen only as indicative, as the
curves are taken at a relatively high bias.

In conclusion, we have introduced a {versatile} 
method to 
deal with strongly correlated
systems out of equilibrium within dynamical mean-field theory. 
The DMFT self-consistent bath is approximated
by an effective one consisting of a small number of sites coupled to
a Markovian bath environment.
The steady state and Green's function of the effective system is
solved by ED of the corresponding Lindblad equation.
The approach is particularly  appropriate to deal directly with the
steady state, without the need to consider full time
evolution. Nevertheless, it should be straightforward, although computationally
more demanding, to deal with  time-dependent problems{, e.g.,} to
describe 
pump-probe processes.

The accuracy of the effective bath to reproduce the DMFT one obviously
depends on the number of bath sites, which is limited by the
exponential increase of the ``super''-Hilbert space. 
Improvements can possibly go along
 solving the Lindblad problem  in the ``smaller'' ordinary
fermion space in combination with quantum trajectory 
methods~\cite{ga.zo,carmichael2,da.ta.09},
and/or by density matrix renormalization group
 approaches~\cite{whit.92.dm,ve.ga.04,pr.zn.09}.

The approach  illustrated here for a simple  {but experimentally relevant~\cite{guenon_electrical_2012}} model
can be extended straightforwardly
to a number of {other} physically relevant systems, including multi layer semiconducting
heterostructures, 
ultracold atoms and correlated
coupled-cavity arrays featuring driving and dissipation, molecular
contacts,  and
can be used to study nonequilibrium quantum phase transitions in these
systems.
Extension to a nonlocal self energy, as in cluster DMFT,
or in nonequilibrium 
 variational/perturbative cluster
approaches ~\cite{kn.li.11,pott.03,ba.po.11} is
also an interesting development.

We acknowledge illuminating discussions with S. Diehl. This work is
partly supported by the Austrian Science Fund (FWF) 
F4103-N13,  J3361-N20,  and 
P24081-N16.



\begin{thebibliography}{73}
\bibitem[{e00()}]{e00too}
\bibinfo{note}{$E_{0,0}$, in general, too, but in the present case it is fixed
  by particle-hole symmetry}.
\bibitem[{alt()}]{alt}
\bibinfo{note}{Alternatively and equivalently, the problem can be solved by the
  ``third quantization'' approach of Ref.~\cite{pros.08}}.
\bibitem[{det()}]{details}
\bibinfo{note}{Details will be given elsewhere}.
\bibitem[{rou()}]{roug}
\bibinfo{note}{Roughly speaking, this is due to the fact that each parameter
  $E_{n,m}$ of the bare system becomes effectively complex and we can use its
  real and imaginary part for the fit}.
\bibitem[{par()}]{para}
\bibinfo{note}{We must fit two functions though, $\Delta^R$ and $\Delta^K$.
  Still the number of parameters per function is twice as large}.
\bibitem[{not()}]{not}
\bibinfo{note}{These are the physical parameters, not to be confused with the
  self-consistent parameters $E$ and $\Gamma$}.
\end{thebibliography}


\section{Supplemental material}

In this supplementary material we present details for the solution of
the effective bath problem described by the Lindblad equation
\beq
\label{li}
\mdot \rho = \li \rho\;,
\eeq
 and the
evaluation of the Green's functions.

\subsection{Mapping onto a ``super'' Fock space}

Following, e.g. \cite{dz.ko.11}, we express the Lindblad operator
$\li=\li_0+\li_I$ 
within a ``super'' Hilbert space
${\cal H}_{sup}\equiv{\cal H}\otimes\wtil{{\cal H}}$
given by the tensor
product of the normal ${\cal H}$ and a ``tilde'' counterpart $\wtil{{\cal H}}$.
Accordingly, one introduces
 fermionic operators $c_n$, as well as their ``tilde'' counterparts 
$\wtil c_n$. 
Within a matrix-vector notation whereby 
\[
\vv c = \left( \begin{array}{c} c_{-N_b} \\ \vdots \\ c_{N_b} \end{array}
\right) \;,
\]
and similarly for $\vvwtil c$ (for simplicity of notation, the spin index
is omitted).
The $\vv c^\dag= (\vv c)^\dag$ are row vectors, and $\vv \Gamma_{1/2},
\vv E$
are the dissipation and ``hopping'' matrices introduced in the text.
The noninteracting part can be expressed~\cite{dz.ko.11} as
\begin{align}
i\li_0=& \sum_{\sigma}\Bigl[ \vv c^{\dag}\left(\vv E-i(\vv \Gamma_{1}-\vv
  \Gamma_{2})\right)\vv c-\vvwtil
 c^{\dag}\left(\vv E+i(\vv \Gamma_{1}-\vv \Gamma_{2})\right)\vvwtil c
\nonumber \\ & \nonumber 
- 2\left(\vvwtil
  c^{T}\vv \Gamma_{1}\vv c+\vvwtil c^{\dag}\vv \Gamma_{2}\vv c^{\dag T}\right)-2\:
i\:\tr\vv \Gamma_{2} \Bigr]\;,
\end{align}
 where $^{T}$ denotes transpose.
The contribution from the interaction is expressed as
\[
i\li_I=
U c_{0\up}^{\dag}c_{0\up}^\nag c_{0\dw}^{\dag}c_{0\dw}^\nag -
U \wtil c_{0\up}^{\dag}\wtil c_{0\up}^\nag \wtil c_{0\dw}^{\dag}\wtil c_{0\dw}^\nag, 
\]

\subsection{Steady state and expectation values}

$\li$, which plays the role of a non-hermitian ``super'' Hamiltonian,
can be diagonalized 
within the many-body Fock-space
${\cal H}_{sup}$
 with conventional methods.
The time dependence of the density matrix $\ket{\rho}$, which is represented as a
Fock state in ${\cal H}_{sup}$, is then expressed
 in terms of the (left) eigenstates $\bra{\alpha L}$
 of $\li$ with eigenvalue $\li_{\alpha}$ as

\begin{equation}
\braket{\alpha L|\rho(t)}=e^{t\li_{\alpha}}\braket{\alpha L|\rho(0)}\label{rhotl}
\end{equation}

Since $i\li$ is non hermitian, $\li_{\alpha}$ have real parts, which
can be shown to be all negative (or zero) , so they produce decaying
behavior in \ref{rhotl}. (At least) one eigenvalue $\li_{\alpha_{0}}=0$.
The corresponding eigenvector $\ket{\alpha_{0}R}$ is the steady state
density matrix.

Following \cite{dz.ko.11},
expectation values and traces are expressed as
\[
\langle\hat{O}\rangle=\tr\hat{O}\rho=\bra I\hat{O}\ket{\rho}=\bra I\hat{O}\rho\ket I
\]
where $\ket I$ is the so called ``left-vacuum state''~\cite{dz.ko.11}. 
Notice that since $0=\tr\mdot{\rho}$, one has $\bra I\li\ket{\rho}=0$ for any
$\ket{\rho}$. Thus, if the steady state is unique, one should have 
\[
\bra I=\bra{\alpha_{0}L}\label{ial}
\]
.

Conservation laws, as usual, reduce the size of the many-body Hilbert space one has to
diagonalize.
For additive conserved quantities, for example $n_{\sigma}$ ($\sigma=\up,\down$),
the density matrix ``state'' is characterized by the constraint
$n_{\sigma}-\wtil n_{\sigma}=0$. In general, one can define ``sectors''
with a given $n_{\sigma}-\wtil n_{\sigma}$, which can be separately
diagonalized. For the Green's functions, these are characterized
by $n_{\sigma}-\wtil n_{\sigma}=\pm 1$

\subsection{Green's function}

To evaluate the Green's function, 
one  needs two time evolutions, i.e. expectation values of the form~\cite{carmichael1}
\beq
\label{gba}
G(B(t),A(t'))\equiv\tr_{U}B(t)A(t')\rho_U=\tr B\; A_{t'}(t-t')
\eeq
for $t>t'$, 
where $\rho_U$ is the density matrix of the universe
(in contrast to $\rho$, which is the reduced density matrix of the
``system''), and $\tr_{U}=\tr \otimes \tr_E$ is the trace over the universe degrees of
freedom, whereby $\tr$ is the one over the system, and $\tr_E$ the one
over the environment.
Here,
\[
A_{t'}(s)\equiv\tr_{E}e^{-isH_U }A \rho_{U}(t')e^{isH_U } \,
\] 
and $H_U$ is the
total Hamiltonian of the universe.

The {\it Quantum regression theorem}~\cite{carmichael1}
 states that under the same
assumptions
for which \eqref{li} holds, one has (for $s>0$)
\beq
\frac{d}{ds}{A_{t'}(s)}=\li A_{t'}(s) \;.
\label{qreg}
\eeq
In terms of eigenvalues and eigenvectors, this gives
\[
\braket{\alpha L|A_{t'}(s)}=e^{s\li_{\alpha}}\braket{\alpha L|A_{t'}(0)}.
\]

For example, the ``greater'' Green's function 
\[
iG_{>n,m}(t,t')\equiv\tr_{U}c_n(t)c_m^{\dag}(t')\rho_U=
\bra Ic_n\: c^\dag_{m t'}(t-t')\ket I
\]
is obtained from 
\eqref{gba}
with $B=c_n$, and $A=c^\dag_m$.
 Inserting the identity $\sum_{\alpha}\ket{\alpha R}\bra{\alpha L}$
and
taking the steady state 
$\ket{\rho(t'=+\infty)}=\ket{\alpha_{0}R}$
this gives
\begin{align*}
iG_{>n,m}&(t'+s,t')= 
\\ &
=\sum_{\alpha}e^{(t-t')\li_{\alpha}}\bra{\alpha_{0}L}c_{n}\ket{\alpha R}\bra{\alpha L}c_{m}^{\dag}\ket{\alpha_{0}R}
\end{align*}
for $t>t'$.
In a similar way, one obtains the result for $t<t'$ and for the
``lesser'' Green's function. 
These are  combined (and Fourier transformed), to
obtain the retarded
($G^R(\omega)$), advanced
$G^A(\omega)$, and Keldysh $G^K(\omega)$ components of the 
Green's function of the effective bath problem evaluated at the
impurity site ($n=m=0$).

The self-energy is extracted by using Dyson's equation in the
$2\times2$ Keldysh space:
\[
\und \Sigma(\omega) = \und G_0(\omega)^{-1} - \und G(\omega)^{-1}  \;.
\]
Using a modified version of 
the ``Langreth rules''~\cite{ha.ja},  one can identify the separate components
\[
\Sigma^R(\omega) = 
1/G_0^R(\omega) - 1/G^R(\omega) \;,
\]
$\Sigma^A(\omega) =  \Sigma^R(\omega)^*$,
and
\[
\Sigma^K(\omega) = 
- G_0^K(\omega)/|G_0^R(\omega)|^2 +
 G^K(\omega)/|G^R(\omega)|^2 \;.
\]

\end{document}